# Spin fluctuations and charge properties of core–shell $C_{80}$+$M_{13}$ (V, Mn, Cr, Ni, Co)

Yixin Li[1], Hongrong Qiu[2], Maolin Bo[1]*

*Corresponding Author: E-mail: bmlwd@yznu.edu.cn (Maolin Bo).

**Abstract**

In this study, we optimized five core–shell structures, $C_{80}$+$M_{13}$ (V, Mn, Cr, Ni, Co), and calculated their electromagnetic properties using density functional theory. We determined that there is electron transfer between $C_{80}$ and the transition metal clusters near the Fermi surface, and that the *d* orbitals contribute most to the magnetism of the structure. $C_{80}Ni_{13}$ was antiferromagnetic. The magnetic properties of clusters have changed significantly, showing Antiferromagnetism. The results establish a theoretical starting point for tuning the electronic and magnetic properties of 13-atom clusters embedded in fullerene cages. Meanwhile, we rotated the Rubik's Cube matrix to obtain a new Hamiltonian expression.

**Key words**
Density functional theory, Transition metal clusters, Core–shell structure, $C_{80}$ clusters

[1]Key Laboratory of Extraordinary Bond Engineering and Advanced Materials Technology (EBEAM) of Chongqing, Yangtze Normal University, Chongqing 408100, China
[2]School of Physics and Optoelectronics, Xiangtan University, Hunan, 411105, China

# 1. Introduction

Clusters are submicroscopic or microscopic aggregates of many atoms, molecules, or ions formed through physical or chemical interactions[1-5]. They range in size from a few angstroms to a few hundred angstroms. In 1962, Kubo et al.[6] found that although bulk metals have continuous energy bands, in metal clusters, these begin to split. In fact, the energy level gap increases with a reduction in cluster size. Thus, clusters manifest certain non-metallic properties with altered specific heat capacity[7]. The properties of the cluster are different from those of a single-atom and bulk material[8-10].

Small clusters of Fe, Co, and Ni have recently been the focus of research owing to their magnetic properties [11, 12]. For example, the average atomic magnetic moment of small iron clusters is 3.0 $\mu$B, which is higher than that of bulk iron[13]. Boudjahem et al.[14] observed a similar increase for small nickel clusters with magnetic moments ranging from 0.67 to 1.33 $\mu$B/atom. Billas et al.[15] studied the relationship between the atomic number and average magnetic moment of atoms for Fe, Co, and Ni clusters and established a relationship curve at 120 K. Here, the average atomic magnetic moment of the Fe clusters was larger than that of bulk iron. Similarly, Apsel et al.[16] studied the magnetic properties of nickel clusters (5–740 atoms) using the Stern–Gallach experiment, determining that smaller clusters have stronger magnetic moments. Furthermore, the magnetic moments of $Fe_{13}$ and $Co_{13}$ are greater in the $I_h$ structures than in the $D_{3h}$ structures, but $Ni_{13}$ appears abnormal[17, 18].

The unique magnetic properties of transition metal clusters have a broad spectrum of future applications in electronic and magnetic devices, which are currently limited by their chemical instability in the environment[19]. The use of fullerene as a protective shell can improve the stability of transition metal clusters. $I_h$-$C_{80}$ is the smallest icosahedral symmetric fullerene and the most unstable of the seven isomer structures of $C_{80}$ that meet the independent pentagon rule. Embedding transition metal clusters into $C_{80}$ can improve their stability via charge transfer from the internal metal atoms to the carbon cage, with $C_{80}$ acquiring a stable closed-shell electronic structure.

Current research has focused on embedding lanthanide monomers or dimers into $C_{80}$. Suzuki et al.[20] embedded $La_2$ into $C_{80}$ and determined that $C_{80}@La_2$ had a higher total charge transfer than monometallic fullerene and a very narrow HOMO–LUMO gap. Furthermore, Svitova et al.[21] prepared the first metal fullerene ($TiLu_2C@I_h\text{-}C_{80}$) and reported several types of bonding in the icosahedral carbon cages. However, there are no reports of 13-atom transition metal clusters embedded in $C_{80}$.

In this study, we constructed a computational model of fullerene $C_{80}$ and embedded transition metal clusters $M_{13}$ (V, Mn, Cr, Ni, Co) to form a core–shell structure (**Fig. 1**). We first constructed and optimized the required structures and determined their theoretical properties, including charge and spin density using DFT calculations and molecular dynamics simulations.

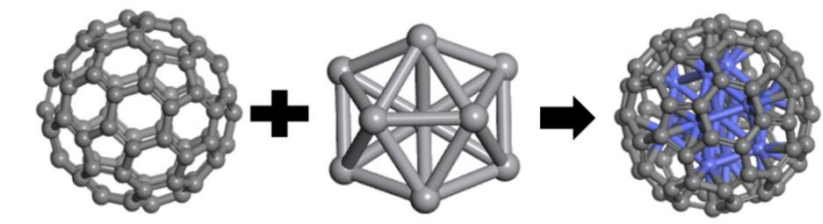

**Fig. 1** Embedding transition metal clusters into $C_{80}$.

## 2. Density functional theory calculations

We simulated core–shell structures of $C_{80}$ embedded with five 13-atom clusters of different transition metals (V, Mn, Cr, Ni, and Co), and used Dmol3 software to optimize the structures and calculate their electronic and magnetic properties. Within the Dmol3 software, the Perdew–Burke–Ernzerhof functional[22] was used to describe electron exchange and correlation potential. Relativistic DFT calculations were conducted using DMol3 software with a double-numeric plus polarization basis set. The core electrons were treated using DFT semi-core pseudopotential. The Brillouin zone $k$ point was set as 2×2×2. An energy convergence of $1.0\times10^{-6}$ eV/atom was considered to obtain a stable structure. All calculations were performed using spin-polarized code.

## 3. Results and discussion

### 3.1 Geometric structure of $C_{80}+M_{13}$ clusters

$C_{80}+M_{13}$ (V, Mn, Cr, Ni, Co) core–shell structures were obtained using Materials Studio software and geometrically optimized by Dmol3. Their lattice parameters after optimization were a = 18.398 Å, b = 18.278 Å, and c = 18.261 Å. The $C_{80}+M_{13}$ optimized core–shell structure is shown in **Fig. 2**, where the superficial carbon layer encases the metal cluster. This fullerene $C_{80}$ layer contains 12 regular pentagons and 30 regular hexagons composed of alternating carbon–carbon double and single bonds. The internal transition metal atoms were bonded to each other and to the fullerene.

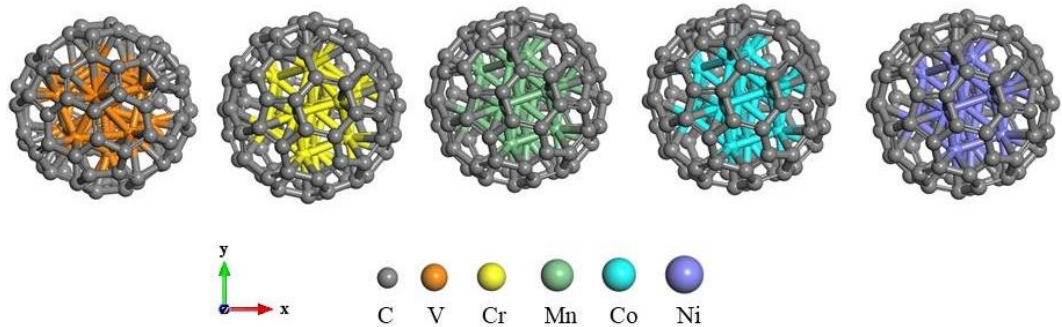

**Fig. 2** Geometric structure of $C_{80}+M_{13}$ clusters. Color coded by atom: gray - C, orange - V, yellow - Cr, green - Mn, blue - Co, and purple - Ni.

We also performed 100 ps molecular dynamics simulations for the five core–shell structures and determined that their total energy remained stable.

### 3.2 Electronic characteristics of $C_{80}+M_{13}$ clusters

The density of states (DOS) provides a way to visualize the energy band structure by characterizing the energy states of all electrons in the system. **Fig. 3** shows the DOS of the five $C_{80}+M_{13}$ clusters, represented as the total DOS (in black), DOS of the transition metal cluster in $C_{80}+M_{13}$ (in red), DOS of $C_{80}$ in $C_{80}+M_{13}$ (in blue), and DOS of individual $C_{80}$ (in pink). **Fig. 3a** indicates that the single $C_{80}$ cluster has the most distinct DOS curve, with sharp peaks possibly caused by electron localization. Furthermore, this cluster has the highest DOS on the Fermi surface, followed by $C_{80}V_{13}$. The lowest DOS was attributed to $C_{80}$ in the core–shell structure.

Thus, the electrons on $C_{80}$ are transferred to the transition metal cluster $V_{13}$, leading to a higher DOS. This can be observed in the range 0.5–1.5 eV, where the DOS of $C_{80}$ clusters is zero; however, the DOS of $C_{80}$ in the core–shell structure is higher due to the transfer of electrons from $V_{13}$ to $C_{80}$.

**Fig. 3b–e** show electron transfer phenomena similar to the other four core–shell structures.

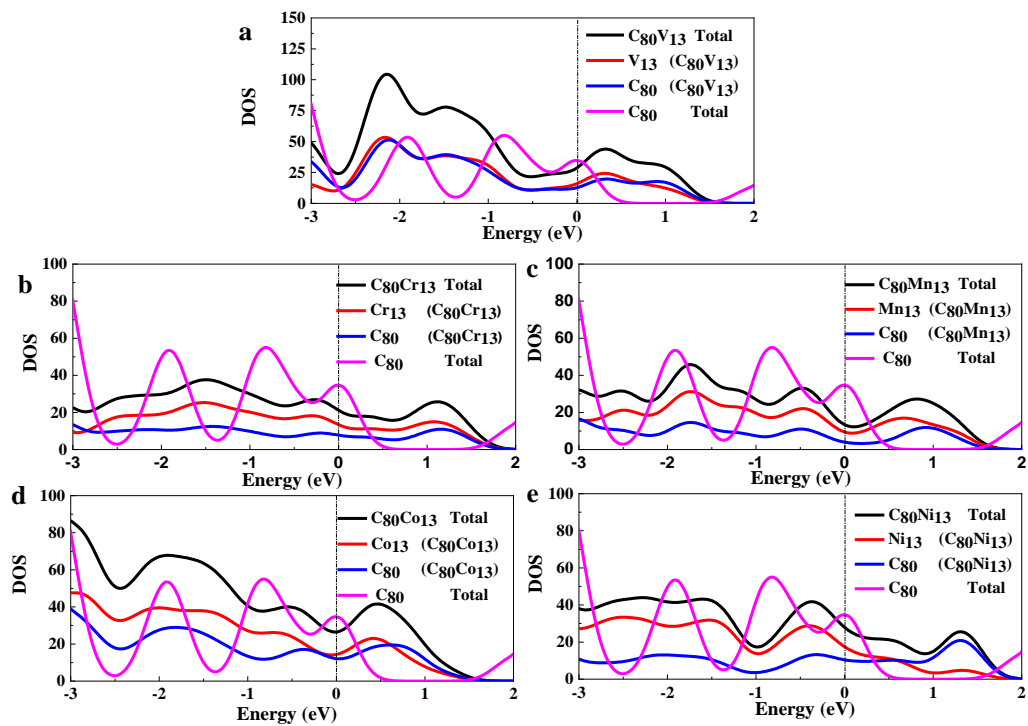

**Fig. 3** Density of states of the $C_{80}+M_{13}$ (V, Cr, Mn, Co Ni) clusters.

### 3.3 Magnetic and spin properties of $C_{80}+M_{13}$ clusters

**Fig. 4** shows the density of spin states diagram of $C_{80}+M_{13}$ clusters with spin-up (in black) and spin-down (in red) curves. Because all the spin-up and spin-down curves have asymmetric characteristics, we can estimate that these structures are magnetic. This asymmetry is most prominent in $C_{80}Co_{13}$. Furthermore, the non-zero spin-up and spin-down DOS of the five core–shell structures in the Fermi surface indicate that these structures are metallic.

The calculated total magnetic moments of $C_{80}V_{13}$, $C_{80}Mn_{13}$, $C_{80}Cr_{13}$, $C_{80}Co_{13}$, and $C_{80}Ni_{13}$, are 0.900, 1.000, 1.968, 9.100, and –3.212 $\mu$B, respectively. Chaves et al.[23] studied the total magnetic moment of 13-atom Au and Pt clusters and determined that the contraction of the *s* orbital and indirect instability of the *d* orbital increased the *s-d* hybridization of the metals and the bond anisotropy and directionality, leading to an open-shell structure. The above results show that the magnetic properties of transition metal clusters can be regulated by constructing $C_{80}+M_{13}$ structures to regulate electronic magnetism.

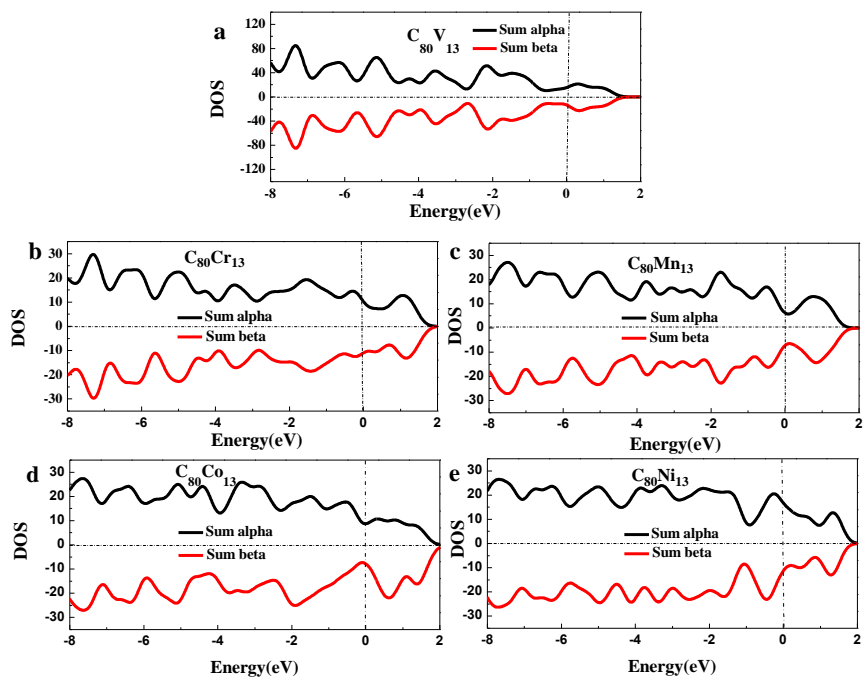

**Fig. 4** Density of spin states of $C_{80}+M_{13}$ (V, Cr, Mn, Co, Ni).

**Fig. 5** shows the density of spin states of the *s* (in black), *p* (in red), and *d* (in blue) orbitals. For C80V13 (**Fig. 5a**), the density of spin states of the *d* orbital was higher than that of the *p* orbital near the Fermi plane. The largest and smallest contributors to spin fluctuation are the *d* and *s* orbitals, respectively. This result was observed for all the energy regions and structures.

Nonetheless, the magnitude of spin fluctuation changed between clusters: $C_{80}V_{13}$ from −2.5 to −2 eV, $C_{80}Cr_{13}$ from −0.5 to 0.5 eV, $C_{80}Mn_{13}$ from −0.5 to 0 eV, $C_{80}Co_{13}$ from −4.5 eV to −4 eV, and $C_{80}Ni_{13}$ from −0.3 to 0 eV. In addition, the density of spin states of $C_{80}Cr_{13}$ and $C_{80}Co_{13}$ were zero according to the Fermi surface theory.

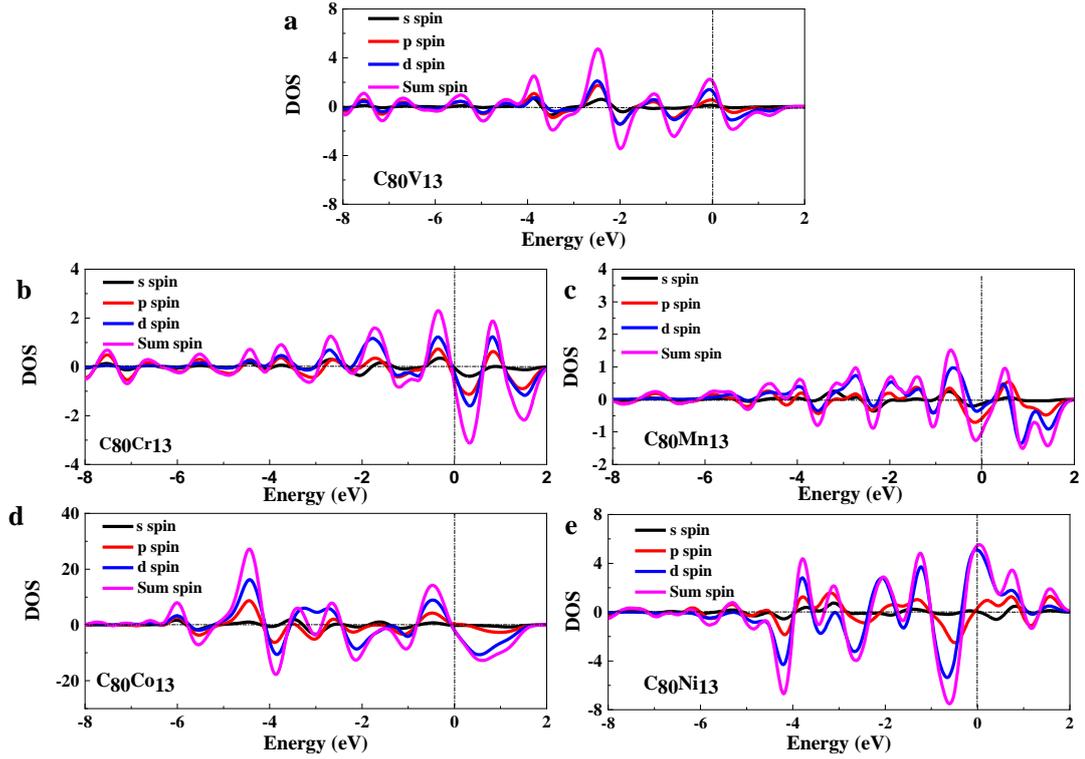

**Fig. 5** Spin DOS of $C_{80}+M_{13}$ (V, Cr, Mn, Co, Ni) clusters

We analyzed the contribution of each atom to charge and spin fluctuations, as shown in **Fig. 6**; the black and red curves represent charge and spin changes, respectively. In **Fig. 6a**, the carbon atoms (atomic positions 1–80) present subtle spin changes, with alternating spin-up and spin-down states. However, vanadium atoms (atomic positions 81–93) present a constant spin-up state with large variations between adjacent atoms. The spin fluctuation of $C_{80}+M_{13}$ was mainly caused by the transition metal clusters. A previous report on the Kondo effect[24] indicated that introducing magnetic impurities significantly contributes to spin fluctuations, which is consistent with our results.

The other four core–shell structures showed similar contributions of $C_{80}$ and the transition metal cluster to spin fluctuation. However, the $Co_{13}$ atoms showed a spin-up state, while the $Ni_{13}$ atoms showed a spin-down state, affecting the magnetism of the core–shell structure. The charge fluctuations were also dominated by changes in the transition metal cluster, with relatively uniform charge distributions for $C_{80}Mn_{13}$, $C_{80}Co_{13}$, and $C_{80}Ni_{13}$.

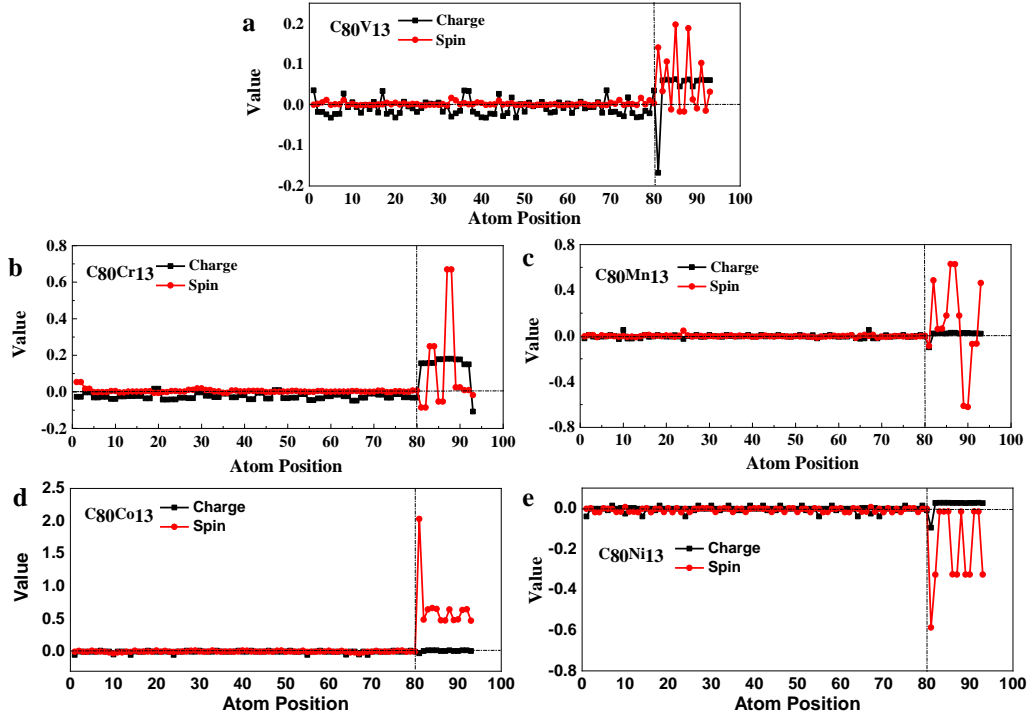

**Fig. 6** Relationship between variable valence elements and spin and charge values.

### 3.4 magic cube matrix (Rubik's cube) operation of spin

The interaction between adjacent atoms in Hamiltonian can be expressed as

$$H = -J_{ij}\vec{S}_i \cdot \vec{S}_j \tag{1}$$

where $J_{ij}$ is the coupling strength.

$$\hat{S}_x = \frac{\hbar}{2}\begin{pmatrix} 0 & 1 \\ 1 & 0 \end{pmatrix}, \quad \hat{S}_y = \frac{\hbar}{2}\begin{pmatrix} 0 & -i \\ i & 0 \end{pmatrix}, \quad \hat{S}_z = \frac{\hbar}{2}\begin{pmatrix} 1 & 0 \\ 0 & -1 \end{pmatrix} \tag{2}$$

Using the relationship between $\hat{S}$ and $\sigma$, the component matrices, namely Pauli matrices, are obtained:

$$\sigma_x = \begin{pmatrix} 0 & 1 \\ 1 & 0 \end{pmatrix}, \quad \sigma_y = \begin{pmatrix} 0 & -i \\ i & 0 \end{pmatrix}, \quad \sigma_z = \begin{pmatrix} 1 & 0 \\ 0 & -1 \end{pmatrix}. \tag{3}$$

The spin expansion of the magic cube matrix[25] can be written as:

$$M_{c\ 2\times 2} = \begin{pmatrix} 0 & 0 \\ 0 & 0 \end{pmatrix} \begin{pmatrix} -i & -i \\ 1 & -1 \end{pmatrix} \begin{pmatrix} 1 & 1 \\ 1 & 1 \end{pmatrix} \begin{pmatrix} 0 & 0 \\ 0 & 0 \end{pmatrix} \begin{pmatrix} 0 & 0 \\ 0 & 0 \end{pmatrix}$$
$$\begin{pmatrix} -1 & 1 \\ i & i \end{pmatrix}$$

(4)

The spin expansion of the magic cube matrix can get the form of the Pauli matrices of the magic cube by rotating:

$$M_{c\ 2\times 2} = \begin{pmatrix} 0 & 0 \\ 0 & 0 \end{pmatrix} \begin{pmatrix} -i & -i \\ 1 & -1 \end{pmatrix} \begin{pmatrix} 1 & 1 \\ 1 & 1 \end{pmatrix} \begin{pmatrix} 0 & 0 \\ 0 & 0 \end{pmatrix} \begin{pmatrix} 0 & 0 \\ 0 & 0 \end{pmatrix} \Rightarrow \sigma_{x\ 2\times 2} = \begin{pmatrix} 0 & i \\ -i & 0 \end{pmatrix} \begin{pmatrix} 1 & 0 \\ 0 & -1 \end{pmatrix} \begin{pmatrix} 0 & 1 \\ 1 & 0 \end{pmatrix} \begin{pmatrix} 0 & -i \\ i & 0 \end{pmatrix} \begin{pmatrix} 1 & 0 \\ 0 & -1 \end{pmatrix} \begin{pmatrix} 0 & 1 \\ 1 & 0 \end{pmatrix}$$
$$\begin{pmatrix} -1 & 1 \\ i & i \end{pmatrix} \qquad \begin{pmatrix} 1 & 0 \\ 0 & -1 \end{pmatrix}$$

$$M_{c\ 2\times 2} = \begin{pmatrix} 0 & 0 \\ 0 & 0 \end{pmatrix} \begin{pmatrix} -i & -i \\ 1 & -1 \end{pmatrix} \begin{pmatrix} 1 & 1 \\ 1 & 1 \end{pmatrix} \begin{pmatrix} 0 & 0 \\ 0 & 0 \end{pmatrix} \begin{pmatrix} 0 & 0 \\ 0 & 0 \end{pmatrix} \Rightarrow [\sigma_y]_{2\times 2} = \begin{pmatrix} 0 & 1 \\ 1 & 0 \end{pmatrix} \begin{pmatrix} 0 & 1 \\ -1 & 0 \end{pmatrix} \begin{pmatrix} 0 & -i \\ i & 0 \end{pmatrix} \begin{pmatrix} 0 & 1 \\ 1 & 0 \end{pmatrix} \begin{pmatrix} 0 & i \\ -i & 0 \end{pmatrix}$$
$$\begin{pmatrix} -1 & 1 \\ i & i \end{pmatrix} \qquad \begin{pmatrix} 0 & -1 \\ 1 & 0 \end{pmatrix}$$

$$M_{c\ 2\times 2} = \begin{pmatrix} 0 & 0 \\ 0 & 0 \end{pmatrix} \begin{pmatrix} -i & -i \\ 1 & -1 \end{pmatrix} \begin{pmatrix} 1 & 1 \\ 1 & 1 \end{pmatrix} \begin{pmatrix} 0 & 0 \\ 0 & 0 \end{pmatrix} \begin{pmatrix} 0 & 0 \\ 0 & 0 \end{pmatrix} \Rightarrow \sigma_{z\ 2\times 2} = \begin{pmatrix} i & 0 \\ 0 & -i \end{pmatrix} \begin{pmatrix} 0 & 1 \\ 1 & 0 \end{pmatrix} \begin{pmatrix} 1 & 0 \\ 0 & -1 \end{pmatrix} \begin{pmatrix} i & 0 \\ 0 & -i \end{pmatrix} \begin{pmatrix} -1 & 0 \\ 0 & 1 \end{pmatrix}$$
$$\begin{pmatrix} -1 & 1 \\ i & i \end{pmatrix} \qquad \begin{pmatrix} 0 & 1 \\ 1 & 0 \end{pmatrix}$$

(5)

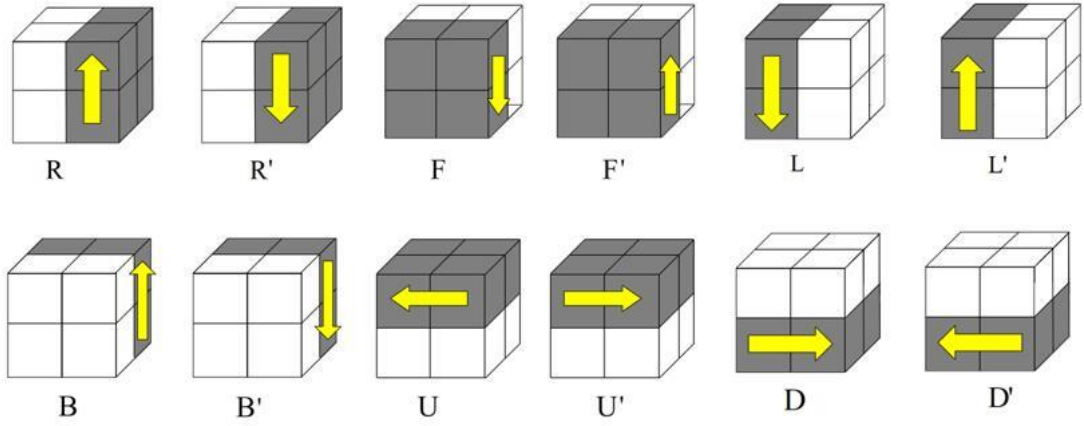

**Fig. 7** Rules for operating the 2th order magic cube matrix

**Fig. 7** shows the operation of the the 2th order magic cube matrix. $F_{m(n)}$, $B_{m(n)}$, $L_{m(n)}$, $R_{m(n)}$, $U_{m(n)}$ and $D_{m(n)}$ are rotation operators of magic cube matrix. $F'_{m(n)}$, $B'_{m(n)}$, $L'_{m(n)}$, $R'_{m(n)}$, $U'_{m(n)}$ and $D'_{m(n)}$ are the rotation operators of the inverse of the magic cube matrix. Transformation of matrix elements through Rubik groups $\Re = \{F_{m(n)}, B_{m(n)}, L_{m(n)}, R_{m(n)}, U_{m(n)}, D_{m(n)}, F'_{m(n)}, B'_{m(n)}, L'_{m(n)}, R'_{m(n)}, U'_{m(n)}, D'_{m(n)}\}$ A letter by itself (e.g. F) means turn that face 90 degrees clockwise with respect to the center of the cube. A letter with an apostrophe (F') denotes a 90 degree counter-clockwise turn. A letter followed by the number 2 ($F_2$) denotes 2 turns, i.e. a 180 degree turn. The subscript n indicates the number of operations. we use the rotation operators of the magic cube matrix $M_c$ to write:

$$\begin{cases} \langle D'_{1(1)} B'_{2(2)} D'_{2(3)} B'_{1(4)} D'_{2(5)} B'_{2(6)} L'_{2(7)} B'_{1(8)} | \ M_{c \ 2\times 2} = \sigma_{x \ 2\times 2} \\ \langle R_{1(1)} U_{1(2)} R'_{1(3)} | \ M_{c \ 2\times 2} = \langle R_{1(1)} U_{1(2)} | \ M_{c \ 2\times 2} | R_{1(3)} \rangle \\ \langle R'_{1(3)} | = | R_{1(3)} \rangle \end{cases}$$

(6)

The magic cube matrix satisfies the rules of matrix operations. For example, the addition operation of the magic cube matrix:

$$\sigma_{x\ 2\times 2}+[\![\sigma_y]\!]_{2\times 2}$$

$$\begin{pmatrix}1&0\\0&-1\end{pmatrix} \qquad \begin{pmatrix}0&1\\-1&0\end{pmatrix}$$

$$=\begin{pmatrix}0&i\\-i&0\end{pmatrix}\begin{pmatrix}0&1\\1&0\end{pmatrix}\begin{pmatrix}0&-i\\i&0\end{pmatrix}\begin{pmatrix}0&1\\1&0\end{pmatrix}+\begin{pmatrix}0&1\\1&0\end{pmatrix}\begin{pmatrix}0&-i\\i&0\end{pmatrix}\begin{pmatrix}0&1\\1&0\end{pmatrix}\begin{pmatrix}0&i\\-i&0\end{pmatrix}$$

$$\begin{pmatrix}1&0\\0&-1\end{pmatrix} \qquad \begin{pmatrix}0&-1\\1&0\end{pmatrix}$$

$$\begin{pmatrix}1&1\\-1&-1\end{pmatrix}$$

$$=\begin{pmatrix}0&i+1\\-i+1&0\end{pmatrix}\begin{pmatrix}0&-i+1\\i+1&0\end{pmatrix}\begin{pmatrix}0&-i+1\\i+1&0\end{pmatrix}\begin{pmatrix}0&i+1\\-i+1&0\end{pmatrix}$$

$$\begin{pmatrix}1&-1\\1&-1\end{pmatrix}$$

.
(7)

For the BBH model[26], we use the 4th order magic cube matrix expansion to calculate the Hamiltonian:

$$\begin{cases} h(k,\delta) = (\gamma + \lambda\cos(k_x))\Gamma_4 + \lambda\sin(k_x)\Gamma_3 + (\gamma + \lambda\cos(k_y))\Gamma_2 + \lambda\sin(k_y)\Gamma_1 + \delta\Gamma_0 \\ \qquad (\Gamma_4) \\ \Gamma_{4\times 4} = (\Gamma_3)(\Gamma_0)(\Gamma_1)(\Gamma_5) \\ \qquad (\Gamma_2) \end{cases}$$

(8)

Here, $\Gamma_0 = \tau_3\sigma_0$, $\Gamma_k = \tau_2\sigma_k$, and $\Gamma_4 = \tau_1\sigma_0$, for $k$ = 1, 2, and 3; $\tau, \sigma$ are Pauli matrices for the degrees of freedom within a unit cell. Among them,

$$\Gamma_0=\tau_3\sigma_0=\begin{vmatrix}1&0&&\\0&1&&\\&&-1&0\\&&0&-1\end{vmatrix}, \Gamma_1=-\tau_2\sigma_1=-\begin{vmatrix}&&0&-i\\&&-i&0\\0&i&&\\i&0&&\end{vmatrix}, \Gamma_2=-\tau_2\sigma_2=-\begin{vmatrix}&&0&-1\\&&1&0\\0&1&&\\-1&0&&\end{vmatrix},$$

$$\Gamma_3=-\tau_2\sigma_3=-\begin{vmatrix}&&-i&0\\&&0&i\\i&0&&\\0&-i&&\end{vmatrix}, \Gamma_4=\tau_1\sigma_0=\begin{vmatrix}&&1&0\\&&0&1\\1&0&&\\0&1&&\end{vmatrix} \text{ and } \Gamma_5=\begin{vmatrix}&&0&0\\&&0&0\\0&0&&\\0&0&&\end{vmatrix}.$$

(9)

So the Hamiltonian expansion is written as:

$$h(k,\delta)=\begin{vmatrix} \delta & 0 & \gamma+\lambda\cos(k_x)+i\lambda\sin(k_x) & \gamma+\lambda\cos(k_y)+i\lambda\sin(k_y) \\ 0 & \delta & -(\gamma+\lambda\cos(k_y))+i\lambda\sin(k_y) & \gamma+\lambda\cos(k_x)-i\lambda\sin(k_x) \\ \gamma+\lambda\cos(k_x)-i\lambda\sin(k_x) & -[\gamma+\lambda\cos(k_y)]-i\lambda\sin(k_y) & -\delta & 0 \\ \gamma+\lambda\cos(k_y)-i\lambda\sin(k_y) & \gamma+\lambda\cos(k_x)+i\lambda\sin(k_x) & 0 & -\delta \end{vmatrix}$$

$$=\begin{vmatrix} \delta & 0 & \gamma+\lambda exp(ik_x) & \gamma+\lambda exp(ik_y) \\ 0 & \delta & -\gamma-\lambda exp(-ik_y) & \gamma+\lambda exp(-ik_x) \\ \gamma+\lambda exp(-ik_x) & -\gamma-\lambda exp(ik_y) & -\delta & 0 \\ \gamma+\lambda exp(-ik_y) & \gamma+\lambda exp(ik_x) & 0 & -\delta \end{vmatrix}$$

(10)

The matrix numerical calculation $(\gamma=1,\lambda=1)$ can obtain its eigenvalues:

$$-e^{-ikx-iky}\sqrt{e^{2ikx+iky}+e^{ikx+2iky}+4e^{2ikx+2iky}+e^{3ikx+2iky}+e^{2ikx+3iky}+e^{2ikx+2iky}\delta^2},$$

$$-e^{-ikx-iky}\sqrt{e^{2ikx+iky}+e^{ikx+2iky}+4e^{2ikx+2iky}+e^{3ikx+2iky}+e^{2ikx+3iky}+e^{2ikx+2iky}\delta^2},$$

$$e^{-ikx-iky}\sqrt{e^{2ikx+iky}+e^{ikx+2iky}+4e^{2ikx+2iky}+e^{3ikx+2iky}+e^{2ikx+3iky}+e^{2ikx+2iky}\delta^2},$$

$$e^{-ikx-iky}\sqrt{e^{2ikx+iky}+e^{ikx+2iky}+4e^{2ikx+2iky}+e^{3ikx+2iky}+e^{2ikx+3iky}+e^{2ikx+2iky}\delta^2}.$$

(11)

The **Fig. 9** is $\delta=1$, the 3D display of the four feature roots of **Eq.10** is shown.

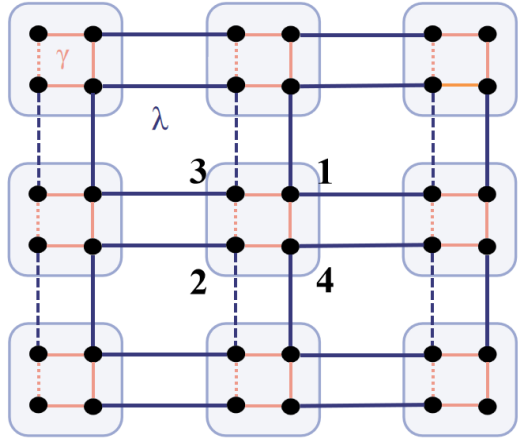

**Fig. 8** $\lambda,\gamma$ represent two hopping strengths, and dashed lines represent hopping terms with negative signs.

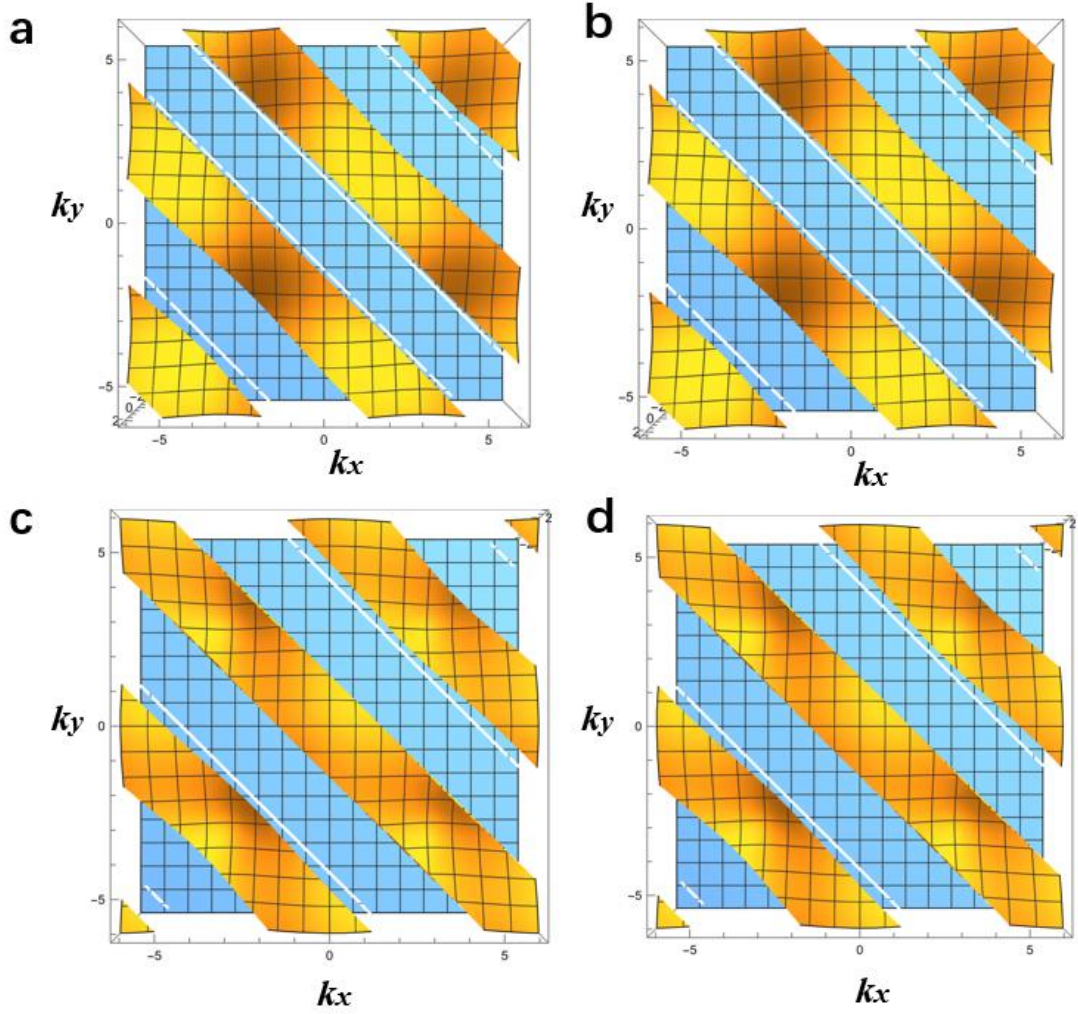

**Fig. 9** the 3D display of the four feature roots of **Eq.10**. (a) (b), (c), and (d) are 1,2,3,4 roots. The cells are (1, 2, 3, 4)，as is shown in the **Fig. 8**. After inverse Fourier transform of the Hamiltonian, we can know that the absolute value of the transition term inside the cell is $\gamma$, and the absolute value of the transition term between cells is $\lambda$. In addition, the transition between 2 and 3 needs to be signed with a negative sign. **Fig. 10** shows the operation of the the 4th order magic cube matrix.

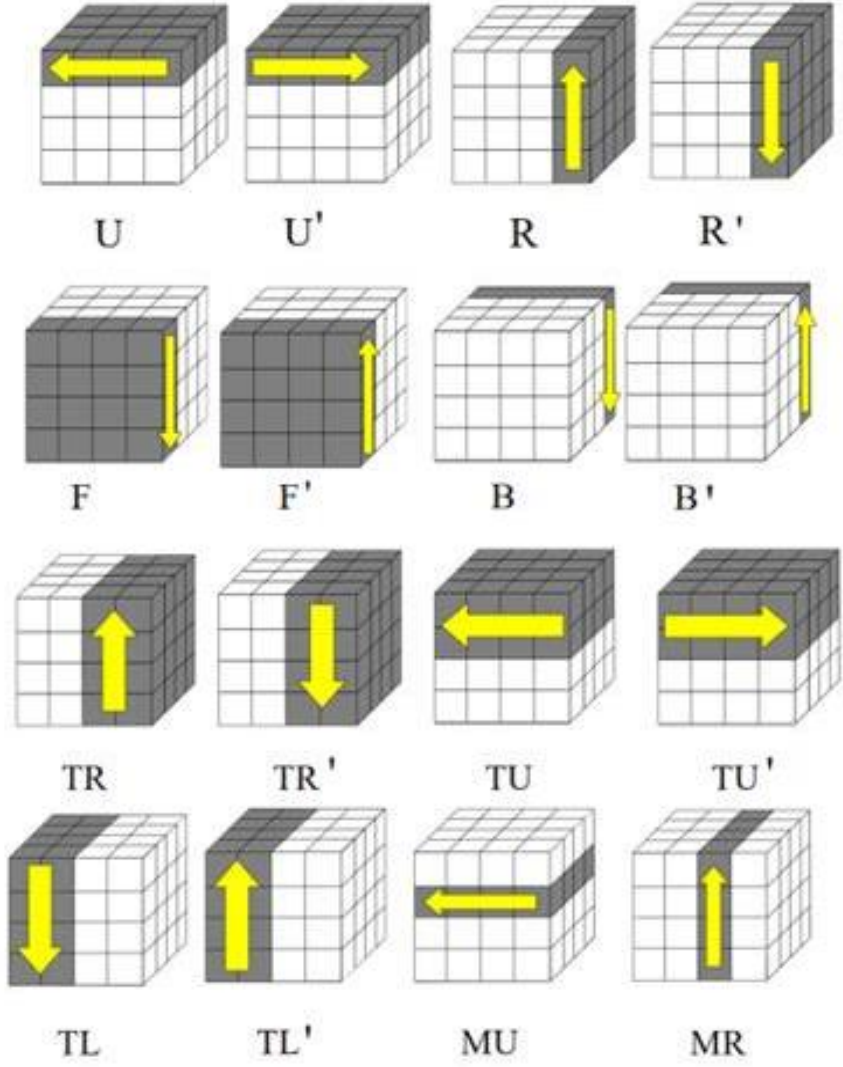

**Fig. 10** Rules for operating the 4th order magic cube matrix

We use the 4th order magic Cube matrix to operate on the BBH Hamiltonian:

$$\begin{cases} h(k,\delta) = (\gamma + \lambda \cos(k_x))\Gamma'_4 + \lambda \sin(k_x)\Gamma'_3 + (\gamma + \lambda \cos(k_y))\Gamma'_2 + \lambda \sin(k_y)\Gamma'_1 + \delta\Gamma'_0 \\ \langle TB_{2(1)} | \ \Gamma_{4\times 4} = \Gamma'_{4\times 4} \\ \qquad (\Gamma'_4) \\ \Gamma'_{4\times 4} = (\Gamma'_3)(\Gamma'_0)(\Gamma'_1)(\Gamma'_5) \\ \qquad (\Gamma'_2) \end{cases}$$

(12)

$$\Gamma'_0 = \tau_3 \sigma_0 = \begin{vmatrix} 1 & 0 & & \\ 0 & 1 & & \\ & & -1 & 0 \\ & & 0 & -1 \end{vmatrix}, \Gamma'_1 = \begin{vmatrix} & & i & 0 \\ & & 0 & -i \\ 0 & -i & & \\ -i & 0 & & \end{vmatrix}, \Gamma'_2 = \begin{vmatrix} & & 0 & 1 \\ & & -1 & 0 \\ 1 & 0 & & \\ 0 & 1 & & \end{vmatrix},$$

$$\Gamma'_3=\begin{vmatrix} & & i & 0 \\ & & 0 & -i \\ i & 0 & & \\ 0 & i & & \end{vmatrix}, \Gamma'_4=\begin{vmatrix} & & 0 & 1 \\ & & -1 & 0 \\ 1 & 0 & & \\ 0 & 1 & & \end{vmatrix} \text{ and } \Gamma'_5=\begin{vmatrix} & & 0 & 0 \\ & & 0 & 0 \\ 0 & 0 & & \\ 0 & 0 & & \end{vmatrix}.$$

(13)

Therefore, the Hamiltonian expansion after operation is written as:

$$h(k,\delta) = (\gamma + \lambda\cos(k_x))\Gamma'_4 + \lambda\sin(k_x)\Gamma'_3 + (\gamma + \lambda\cos(k_y))\Gamma'_2 + \lambda\sin(k_y)\Gamma'_1 + \delta\Gamma'_0$$

$$h(k,\delta)=\begin{vmatrix} \delta & 0 & i\lambda(\sin(k_x)+\sin(k_y)) & 2\gamma+\lambda(\cos(k_x)+\cos(k_y)) \\ 0 & \delta & -2\gamma-\lambda(\cos(k_x)+\cos(k_y)) & -i\lambda(\sin(k_x)+\sin(k_y)) \\ 2\gamma+\lambda(\cos(k_x)+\cos(k_y))+i\lambda\sin(k_x) & -i\lambda\sin(k_y) & -\delta & 0 \\ -i\lambda\sin(k_y) & 2\gamma+\lambda(\cos(k_x)+\cos(k_y))+i\lambda\sin(k_x) & 0 & -\delta \end{vmatrix}$$

$$=\begin{vmatrix} \delta & 0 & i\lambda(\sin(k_x)+\sin(k_y)) & 2\gamma+\lambda(\cos(k_x)+\cos(k_y)) \\ 0 & \delta & -2\gamma-\lambda(\cos(k_x)+\cos(k_y)) & -i\lambda(\sin(k_x)+\sin(k_y)) \\ 2\gamma+\lambda(\cos(k_y)+e^{ik_x}) & -i\lambda\sin(k_y) & -\delta & 0 \\ -i\lambda\sin(k_y) & 2\gamma+(\lambda\cos(k_y)+e^{ik_x}) & 0 & -\delta \end{vmatrix}$$

(14)

The Hamiltonian matrix is constantly changing with the rotation of the Rubik's Cube, and solving the Hamiltonian is solving the Rubik's Cube matrix.

The matrix numerical calculation $(\gamma=1, \lambda=1)$ can obtain its eigenvalues:

$$-e^{-kx}\sqrt{e^{2kx}\delta^2 - \sqrt{2}\sqrt{-e^{2kx}(3+2\cos[kx]+\cos[kx-ky]+2\cos[ky])(1+4e^{kx}+5e^{2kx}+2e^{kx}\cos[ky]+4e^{2kx}\cos[ky])}},$$

$$e^{-kx}\sqrt{e^{2kx}\delta^2 - \sqrt{2}\sqrt{-e^{2x}(3+2\cos[kx]+\cos[kx-ky]+2\cos[ky])(1+4e^{kx}+5e^{2kx}+2e^{kx}\cos[ky]+4e^{2kx}\cos[ky])}},$$

$$-e^{-kx}\sqrt{e^{2kx}\delta^2 + \sqrt{2}\sqrt{-e^{2kx}(3+2\cos[kx]+\cos[kx-ky]+2\cos[ky])(1+4e^{kx}+5e^{2kx}+2e^{kx}\cos[ky]+4e^{2kx}\cos[ky])}},$$

$$e^{-kx}\sqrt{e^{2kx}\delta^2 + \sqrt{2}\sqrt{-e^{2kx}(3+2\cos[kx]+\cos[kx-ky]+2\cos[ky])(1+4e^{kx}+5e^{2kx}+2e^{kx}\cos[ky]+4e^{2kx}\cos[ky])}}.$$

(15)

The **Fig. 11** is $\delta=1$, the 2D display of the four feature roots of **Eq.14** is shown.

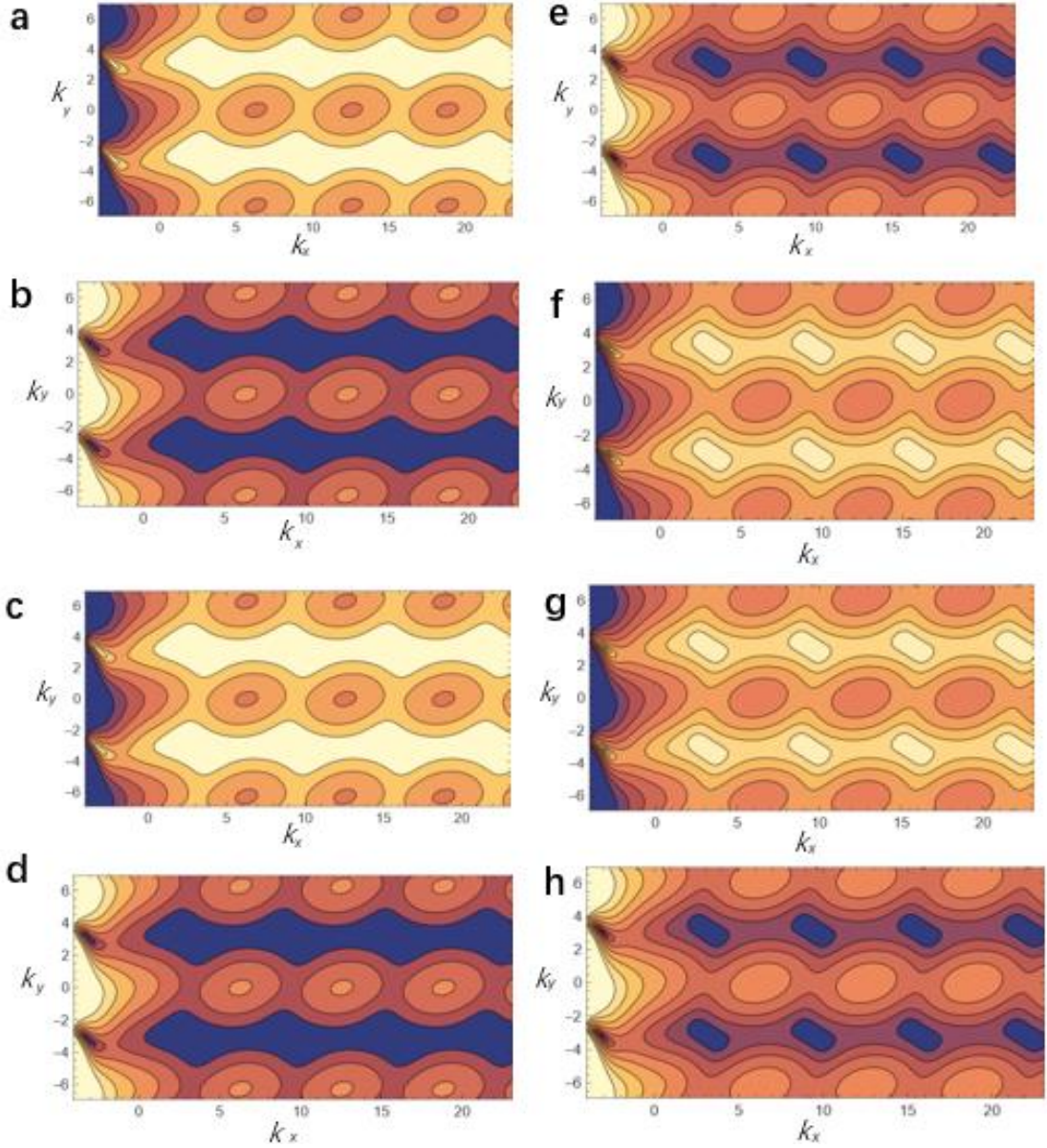

**Fig. 11** the 2D display of the four feature roots of **Eq.14**. (a) (b), (c), and (d) are the real parts of 1,2,3,4 roots. (e), (f), (g), and (h) are the imaginary parts of 1,2,3,4 roots.

**3.5 three-dimensional maps of spin**

**Fig. 12** shows three-dimensional maps of the atomic positions, spin-up values, and spin-down values obtained by the spatial symmetry operation of spin under the assumption that the five closed-shell structures are chiral. However, these graphs cannot be obtained using simple rotations.

The three-dimensional graphs clearly show the ratio of spin-up and spin-down atoms; a magnetic moment is generated by an unbalanced atomic spin. In addition, it reveals which atoms contribute more to the spin fluctuation. In $C_{80}V_{13}$, vanadium atoms 5, 8, 1, 3, and 11 have greater influence on the spin fluctuation. Furthermore, chromium atoms 7, 8, 3, and 4 and cobalt atom 1 have a greater influence on the spin fluctuation of $C_{80}Cr_{13}$ and $C_{80}Co_{13}$, respectively. In these three structures, the atoms with the greatest influence are in the spin-up state. In contrast, the seven atoms with the greatest influence on the spin fluctuation of $C_{80}Ni_{13}$ are in the spin-down state. Finally, two of the seven atoms with the greatest influence on spin fluctuation in $C_{80}Mn_{13}$ are in a spin-down state.

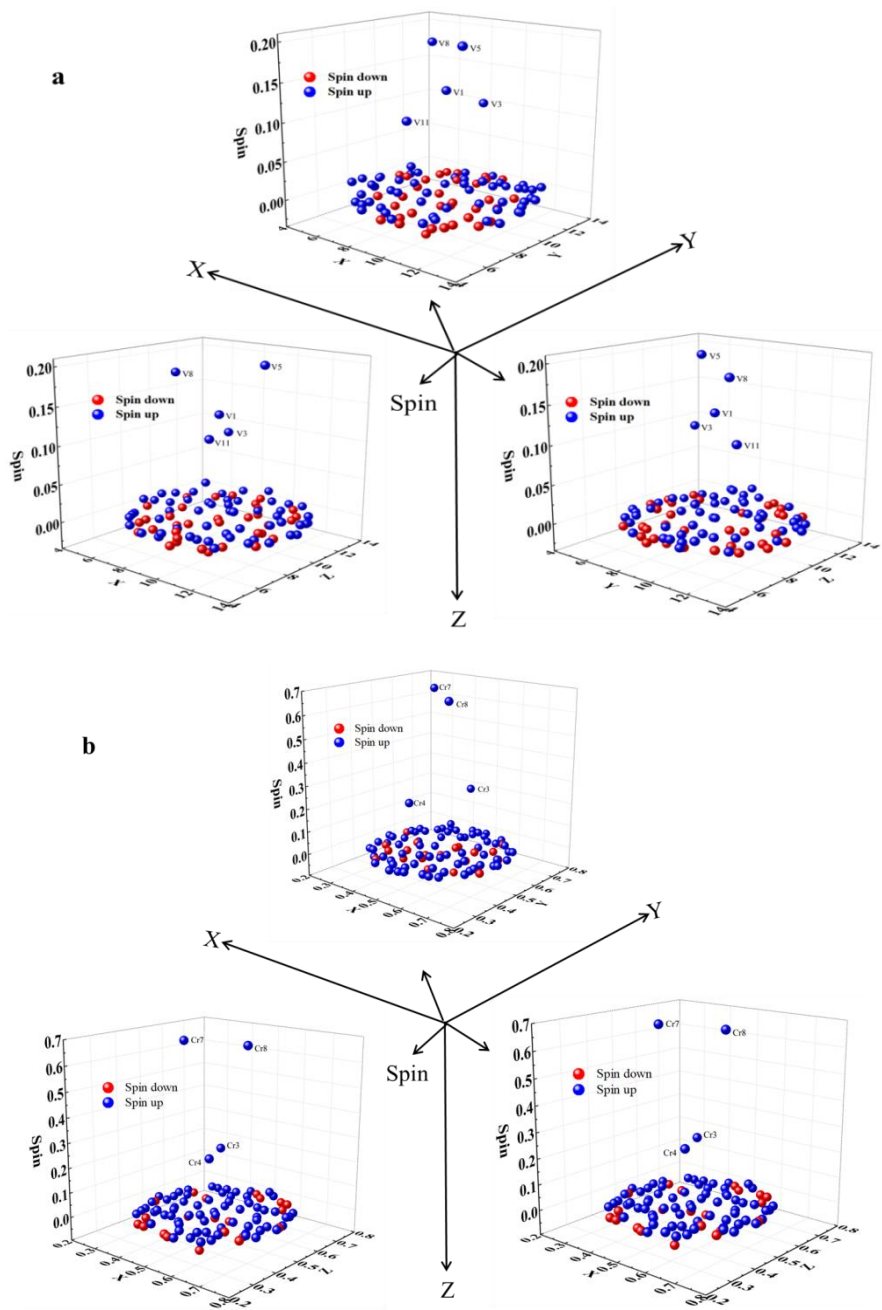

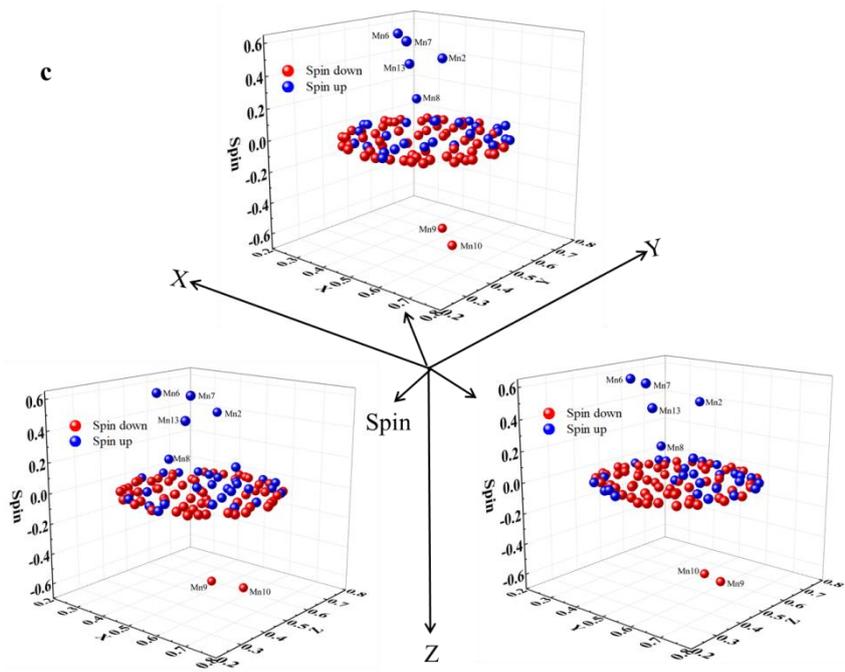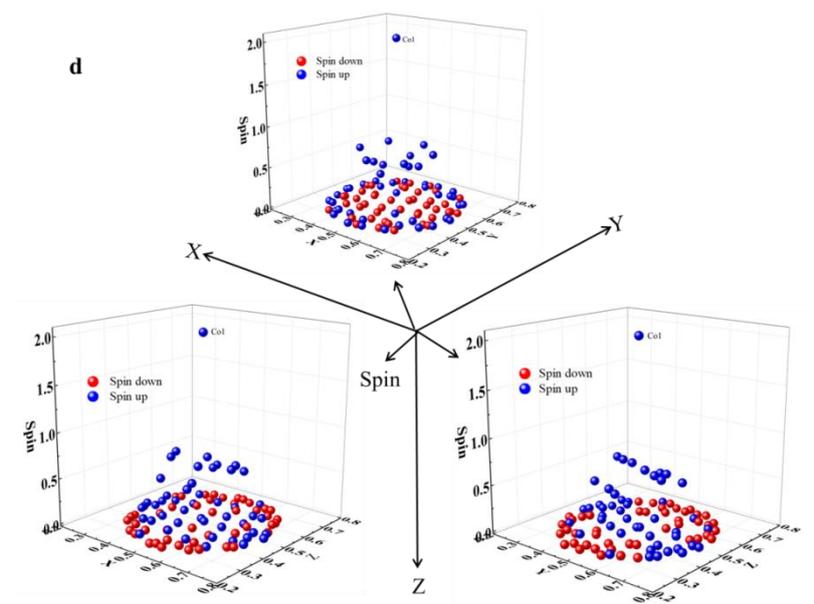18

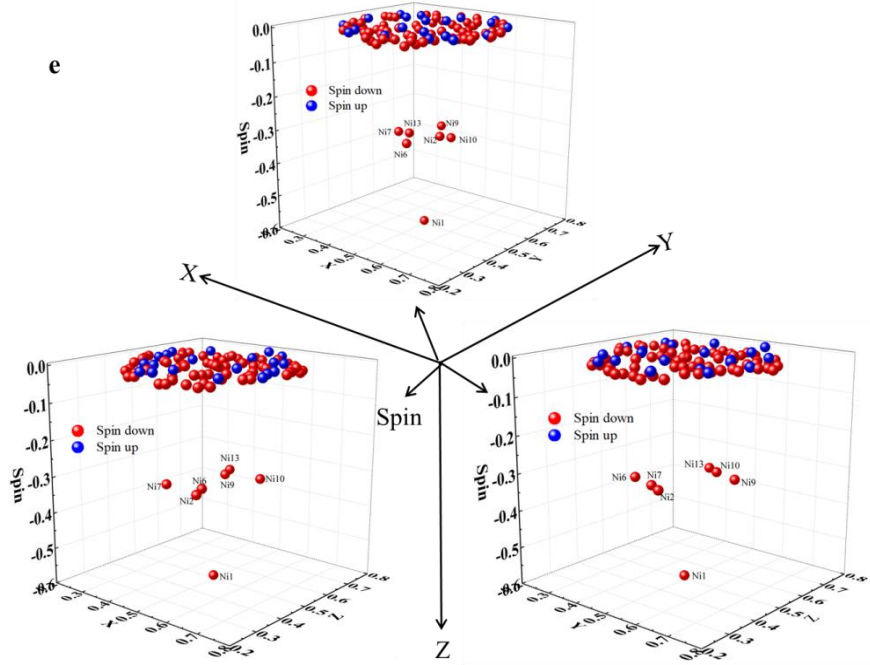

**Fig. 11** Three-dimensional graph of atomic position, spin-up values, and spin-down values for $C_{80}V_{13}$ (a), $C_{80}Cr_{13}$ (b), $C_{80}Mn_{13}$ (c), $C_{80}Co_{13}$ (d), and $C_{80}Ni_{13}$ (e).

## 4. Conclusions

In this study, we simulated five core–shell structures of 13-atom transition metal clusters with unique magnetic properties encased by a $C_{80}$ carbon cage, $C_{80}+M_{13}$ (V, Mn, Cr, Ni, Co), and calculated their electronic and magnetic properties using DFT calculations. The DOS of the five core–shell structures was compared with that of the isolated $C_{80}$ carbon cage, revealing electron transfer between $C_{80}$ and the transition metal clusters near the Fermi surface. In the 0.5–1.5 eV energy range, the DOS of isolated $C_{80}$ and $C_{80}$ in the core–shell structures was zero and non-zero, respectively, indicative of electron transfer from the metal cluster to the carbon cages. The calculated total magnetic moment of the $C_{80}+M_{13}$ structures revealed antiferromagnetic properties that were inconsistent with the magnetic moments of the corresponding 13-atom transition metal cluster. This indicates that the electronic magnetic properties can be controlled by constructing $C_{80}+M_{13}$ structures. Furthermore, the *d* orbitals contributed the most to the magnetic moments of these structures.

Finally, we analyzed the effect of all atoms on the charge and spin fluctuations and determined that the transition metal cluster has the greatest impact. Through the spatial symmetry operation of spin, we created a three-dimensional map of the atomic positions, spin-up, and spin-down, which enabled us to determine the ratio and individual contributions of each atom. This paper provides a theoretical basis for the regulation of the electronic and magnetic properties transition metal clusters by embedding them into fullerene cages. We established the rotation matrix of Hamiltonian is a novel way to operation of spin.